\documentclass{elsart}

\usepackage{amssymb}
\usepackage{graphicx}
\usepackage{verbatim}
\usepackage{longtable}
\usepackage{dcolumn}

\newcommand{\be}{\begin{equation}}
\newcommand{\ee}{\end{equation}}
\newcommand{\bea}{\begin{eqnarray}}
\newcommand{\eea}{\end{eqnarray}}

\newcommand{\cro}[1]{\left[#1\right]}

\newcommand{\avg}[1]{\langle{#1}\rangle}
\newcommand{\ovl}[1]{\overline{#1}}
\newcommand{\BE}{\begin{eqnarray}}
\newcommand{\EE}{\end{eqnarray}}
\newcommand{\BEn}{\begin{eqnarray*}}
\newcommand{\EEn}{\end{eqnarray*}}
\newcommand{\barr}{\begin{array}}
\newcommand{\earr}{\end{array}}

\newcommand{\bit}{\begin{itemize}}
\newcommand{\eit}{\end{itemize}}
\newcommand{\bc}{\begin{center}}
\newcommand{\ec}{\end{center}}
\newcommand{\ben}{\begin{enumerate}}
\newcommand{\een}{\end{enumerate}}

\newcommand{\eps}{\epsilon}

\begin{document}

\begin{frontmatter}



\title{Feedback and efficiency in limit order markets}


\author{Damien Challet}

\address{D\'epartement de Physique, Universit\'e  de Fribourg, 1700 Fribourg, Switzerland, and\\Institute for Scientific Interchange,
Viale S. Severo 65,
10133 Torino, Italy}

\begin{abstract}
A consistency criterion for price impact functions in limit order markets is proposed that prohibits chain arbitrage exploitation.  Both the bid-ask spread and the feedback of sequential market orders of the same kind onto both sides of the order book are essential to ensure consistency at the smallest time scale. All the stocks investigated in Paris Stock Exchange have consistent price impact functions.

\end{abstract}

\end{frontmatter}

\maketitle
\section{Introduction}

Mainstream finance and mathematical finance suppose that the price dynamics follows a random walk \cite{bachelier,Fama}. This is an extreme point of view
describing an average idealised behaviour that does not account for every
detail of the microscopic price dynamics. And indeed extreme assumptions are
most useful in a theoretical framework. This is why the opposite
one  is worth considering \cite{C07}: suppose that trader $0$ is active at time $t$; he buys/sells a
given amount of shares $n_0$,  leading to (log-)price change $r(t)=r_0$, where
$t$ is in transaction time, $t$ being the $t$-th transaction. In addition,
one also assumes that trader $1$ has perfect information about $t$ and
$r_0$ and exploits it accordingly.

A related situation is found in Ref.\ \cite{FarmerForce} whose main result is that a arbitrage opportunity, when exploited, does not disappear but is spread around $t$. It is a counter-intuitive outcome, that raises two questions: how to accommodate the never-disappearing arbitrage, and how microscopic arbitrage removal is possible at all at this time scale. This proceeding, a short version of Ref.\ \cite{C07}, suggests that real markets remove arbitrage on a single transaction basis by a double feedback of the last transactions on the order book.

The price impact function $I(n)$ is by definition the relative price
change caused by a transaction of $n$ (integer) shares ($n>0$ for
buying, $n<0$ for selling); mathematically,

\be
p(t+1)=p(t) + I(n),
\ee

\noindent where $p(t)$ is the {\em log}-price and $t$ is in transaction
time. The above notation misleadingly suggests that $I$ does not depend
on time. In reality, $I$ is not only  subject to random fluctuations
(which will be neglected here), but also, for instance, to
feed-back from the type of market orders which has a long
memory (see e.g. \cite{BouchaudLimit3,FarmerSign,BouchaudLimit4,FarmerMolasses} for
discussions about the dynamical nature of market impact). Neglecting the
dynamics of $I$ requires us to consider specific shapes for $I$ that
enforce some properties of price impact for each transaction, whereas in
reality they only hold on average. For example, one should restrict
oneself to the class of functions that makes it impossible to obtain
round-trip positive gains \cite{FarmerForce}. But the inappropriateness
of constant price impact functions is all the more obvious as soon as
one considers how price predictability is removed by speculation, which
is inter-temporal by nature.

The most intuitive (but wrong) view of market inefficiency is to regard
price predictability as a scalar deviation from the unpredictable case:
if there were a relative price deviation $r_0$ caused by a transaction
of $n_0$ shares at some time $t$, according to this view, one should
exchange $n_1$ shares so as to cancel perfectly this anomaly, where
$n_1$ is such that  $I(n_1)=-r_0$. This view amounts to regarding
predictability as something that can be remedied with a single trade.
However, the people that would try and cancel $r_0$ would not gain
anything by doing it unless they are market makers who try to stabilise
the price.

It is most instructive to understand how constant price impact functions are paradoxical by considering a simple example. Trader $1$, a perfectly (and possibly illegally) informed speculator, will
take advantage of his knowledge by opening a position at time $t-1$ and
closing it at time $t+1$. It is important to be aware that if one places
an order at time $t$, the transaction takes place at price $p(t+1)$. Provided
that trader $0$ buys/sells $n_0$ shares irrespective of the price that
he obtains, the round-trip of trader $1$ yields a monetary gain of $$g_1=n_1[e^{p(t+2)}-e^{p(t)}]=n_1e^{p_0}[e^{I(n_0)}-e^{I(n_1)}]$$
where $p_0$ is the log-price before any trader considered here
makes a transaction. Since $I(n)$ generally increases with $n$, there is
an optimal $n_1^*$ number of shares  that maximises $g_1$.
The discussion so far is a simplification, in real-money instead of
log-money space,  of the one found in Ref. \cite{FarmerForce}. One
should note that far from diminishing price predictability, the
intervention of trader $1$ increases the fluctuations. Therefore, in the
framework of constant price impact functions, an isolated arbitrage
opportunity never vanishes but becomes less and less exploitable because
of the fluctuations, thus the reduction, of signal-to-noise ratio caused
by the speculators.

It seems that trader $1$ cannot achieve a better gain than by holding
$n_1^*$ shares at time $t$. Since the actions of trader $1$ do not
modify in any way the arbitrage opportunity between $t-2$ and $t+2$, he
can inform a fully trusted friend, trader $2$, of the gain opportunity
on the condition that the latter opens his position before $t-1$ and
closes it after $t+1$ so as to avoid modifying the relative gain of
trader $1$.\footnote{If trader $2$ were not a good friend, trader $1$ could in principle ask trader $2$ to open his position after him and to close it after him, thus earning more. But relationships with real friends are supposed to egalitarian in this paper.} For instance, trader $2$ informs trader $1$ when he has
opened his position and trader $1$ tells trader $2$ when he has closed
his position. From the point of view of trader $2$, this is very
reasonable because the resulting action of trader $1$ is to leave the
arbitrage opportunity unchanged to $r_0$ since $p(t+1)-p(t-1)=r_0$.
Trader $2$ will consequently buy $n_2^*=n_1^*$ shares at time $t-2$ and
sell them at time $t+2$, earning the same return as trader $1$. This can
go on until trader $i$ has no fully trusted friend. Note that the
advantage of trader $1$ is that he holds a position over a smaller time
interval, thereby increasing his return rate; in addition, since trader $2$
increases the opening price of trader $1$, the absolute monetary
gain of trader $1$ actually {\em increases} provided that he has enough
capital to invest. Before explaining why this situation is paradoxical, it makes
sense to emphasise that the gains of traders $i>0$ are of course obtained
at the expense of trader $0$, and that the result the particular order
of the traders' actions is to create a bubble which peaks at time $t+1$.

The paradox is the following: if trader $1$ is alone, the best return
that can be extracted from his perfect knowledge is $\hat g_1(n_1^*)$
according to the above reasoning. When there are $N$ traders in the ring
of trust, the total return extracted is $N$ times the optimal gain of a
single trader. Now, assume that trader $1$ has two brokering accounts;
he can use each of his accounts, respecting the order in which
to open and close his positions, effectively earning the optimal return
on each of his accounts. The paradox is that his actions would be
completely equivalent to investing $n_1^*$ and then $n_1^*$ from the
same account. In particular, in the case of $I(n)=n$, this seems {\em a
priori} exactly similar to grouping the two transactions into $2n_1^*$,
but this results of course in a return smaller than the optimal return for
a doubled investment. Hence, in this framework, trader $1$ can earn as
much as  pleases  provided that he splits his investment into
sub-parts of $n_1^*$ shares whatever $I$ is, as long as it is constant.

This paradox seems too good to be present in real markets. As a consequence, one should rather consider its impossibility as an {\em a contrario}  consistency criterion for price impact functions. Let us introduce the two relevant mechanisms that are at work in real markets.

Half of the solution lies in the dynamics of the order book, particularly the reaction of the order book to a sequence of market orders of the same kind. Generically, the impact of a second market order of the same kind and size is smaller by a factor $\kappa_1$ than that of the first one, and similarly by a factor $\kappa_2$ for a third one, etc \cite{BouchaudLimit5,FarmerMolasses}. To this contraction of market impact on one side also corresponds an increase of market impact on the other side for the next market order of opposite type \cite{BouchaudLimit5}; therefore, we shall assume that the impact function on the other side is divided by $\theta_1$ after the first market order, by $\theta_2\theta_1$ after the second, etc. As shown in the next section, $\kappa_1\simeq\kappa_2$ is a very good approximation when $\kappa_1$ and $\kappa_2$ are averaged over all the stocks, hence we shall only use $\kappa$; for the same reason, we assume that $\theta_1=\theta_2=\theta$.

\section{Feedback}

In order to investigate whether the feedback restricted on the side on which the first sequential market orders are placed is enough to make price impact consistent, one sets $\theta=1$. In the case of $\log$ price impact functions, the optimal number of shares and gain of trader $1$ are
\be
n_1^*=\frac{n_0^\kappa}{(\gamma+1)^{1/\gamma}}
\ee
and
\be\label{eq:g1optdyn}
g_1^*=e^{p_0}n_0^{\kappa(\gamma+1)}\frac{\gamma }{(\gamma+1)^{1+1/\gamma}}.
\ee
These two equations already show  that the reaction of the limit order book reduces the gain opportunity of player $1$. Adding trader $2$ will reduce further the impact of trader $0$, hence the gain of trader $1$, and, as before, trader $2$ should pay for it. In this case, the reduction of gain of trader $1$ is
\bea
\frac{\Delta g_1}{e^{p_0}}&=&[g_1^*-g_1(n_1^*,n_2)]e^{-p_0}=n_0^{\kappa(\gamma+1)}\frac{\gamma}{(\gamma+1)^{1+1/\gamma}}\nonumber\\&&-n_2^\gamma\frac{n_0^{\gamma(1+\kappa\gamma)}}{(\gamma+1)^{1/\gamma+\kappa}}[n_0^{-\gamma\kappa(1-\kappa)}(\gamma+1)-1],
\eea
while the gain that trader $2$ optimises is
\be
\frac{G_2}{e^{p_0}}=n_2^{\gamma+1}\cro{\frac{1}{n_2^{\kappa\gamma}}\frac{n_0^{\kappa\gamma(2\kappa-1)}}{(1+\gamma)^{\kappa-1}}-1}-\frac{\Delta  g_1}{e^{p_0}}.
\ee


Trader $1$'s impact functions are $\kappa I$ when he opens his position and $I$ when he closes it, which is an additional cause of loss for trader $1$, which must be also compensated for by trader $2$. Fortunately for the latter, his impact functions are $I$ when opening and $\kappa I$ when closing his position. Therefore, provided that $\kappa$ is large enough so as not to make $\kappa^2 I(n_0)$ too small, trader $2$ can earn more than trader $1$ in some circumstances. Impact functions are inconsistent when $G_2^*>0$, $n_1^*>1$ and $n_2^*>1$ for $\log$ impact functions. It turns out that the regions in which $G_2^*>0$ while $n_1^*>1$ are disjoint if $\kappa<\kappa_c\simeq 0.5$ for $\log$ impact functions.


As previously mentioned, the feedback acts on both order book sides. Assuming that $\theta_1=\theta_2=\kappa$ in order to be able to use available market measurements, one finds a critical value of $\kappa$ of about $0.83$ for $\log$ price impact functions. For this value, only a small area of inconsistent impact functions, corresponding to $n_0\simeq 1.5$, still exists in the $(n_0,\gamma)$ plane, but cannot be reached since both $n_0=1$ and $2$ are outside of the inconsistent region. Therefore, even double feedback does not guarantee consistency

\section{Empirical data}


The values of $\kappa_1$ and $\kappa_2$ can be measured in real markets.
The response function $R(\delta t,V)=\avg{(p(t+\delta t)-p(t))\eps(t)}|_{V(t)=V}$ is the average price change after $\delta t$ trades, conditional on the sign of the trade $\eps(t)$ and on volume $V$; similarly, one defines the response function conditional on two trades of the same sign $R^+(\delta t,V)=\avg{(p(t+\delta t)-p(t))\eps(t)}|_{\eps(t)=\eps(t-1),V(t)=V}$, and $R^{++}(\delta t)=\avg{(p(t+\delta t)-p(t))\eps(t)}|_{\eps(t)=\eps(t-1)=\eps(t-2)},V(t)=V$. A key finding of Refs \cite{BouchaudLimit3,BouchaudLimit4} is that $R$ factorises into $R(\delta t)F (V)$. Thus we will be interested in $R(\delta t)$, $R^+(\delta t)$ and $R^{++}(\delta t)$.

Using measures kindly provided by J.-Ph. Bouchaud and J. Kockelkoren, one finds that the estimate of this ratio $\hat \kappa=\avg{R^+(1)}/\avg{R(1)}\in[0.86,1.02]$, that the average over all stocks $\ovl{\hat\kappa_1}=0.97\pm0.04$ and $\hat \kappa_2=\avg{R^{++}(1)}/\avg{R^+(1)}\in[0.87,1.01]$, while $\ovl{\hat \kappa_2}=0.97\pm0.03$. ; for a given stock, there is some correlation between $\kappa_1$ and $\kappa_2$; the data presented here does not contain error bars for the measures of $R$, $R^+$ and $R^{++}$. The approximation $\kappa_1\simeq \kappa_2$ is reasonable, and we shall from now on call $\kappa=(\kappa_1+\kappa_2)/2$ and replace $\kappa_1$ and $\kappa_2$ by $\kappa$ everywhere.

In other words there is some variations between the stocks, some of them being less sensitive to successive market orders of the same kind. The values of estimated $\kappa$ start at $0.86$. Therefore, even feedback on both book sides does not yield consistent $\log$ impact functions. One concludes that the feedback of the order book is not enough to make price impact functions consistent.

\section{Spread}

\begin{figure}
 \centerline{\includegraphics[width=0.4\textwidth]{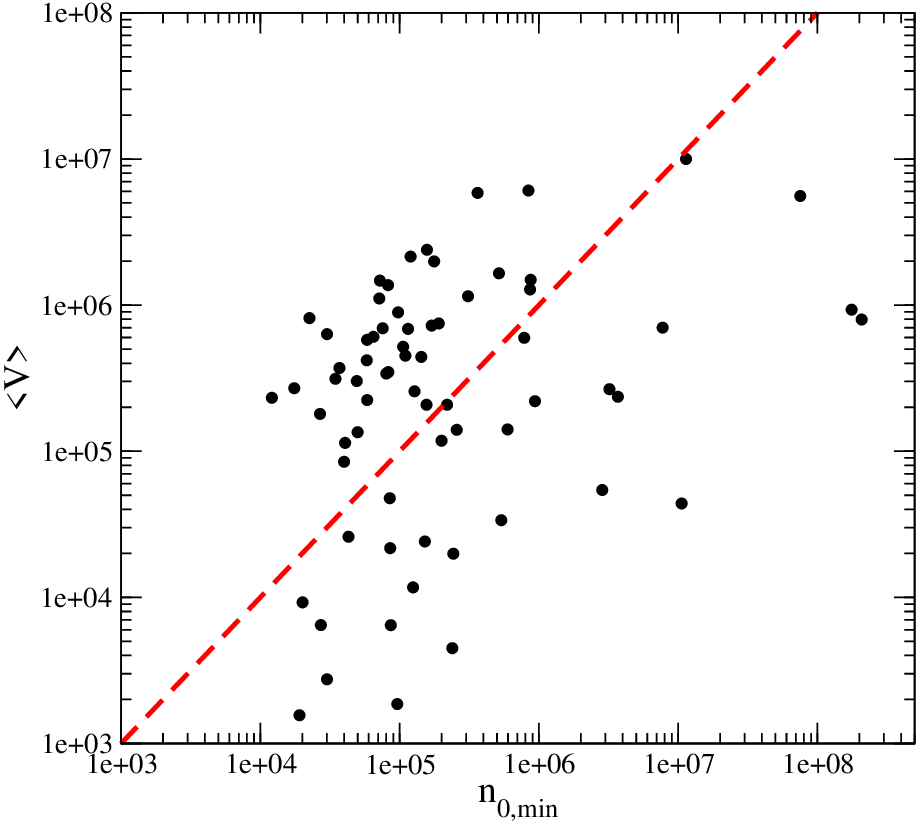}}
\caption{Average daily volume versus $n_{0,min}$ (same data set)}\label{fig:avgV_vs_n0min}
\end{figure}

\begin{figure}
 \centerline{\includegraphics[width=0.4\textwidth]{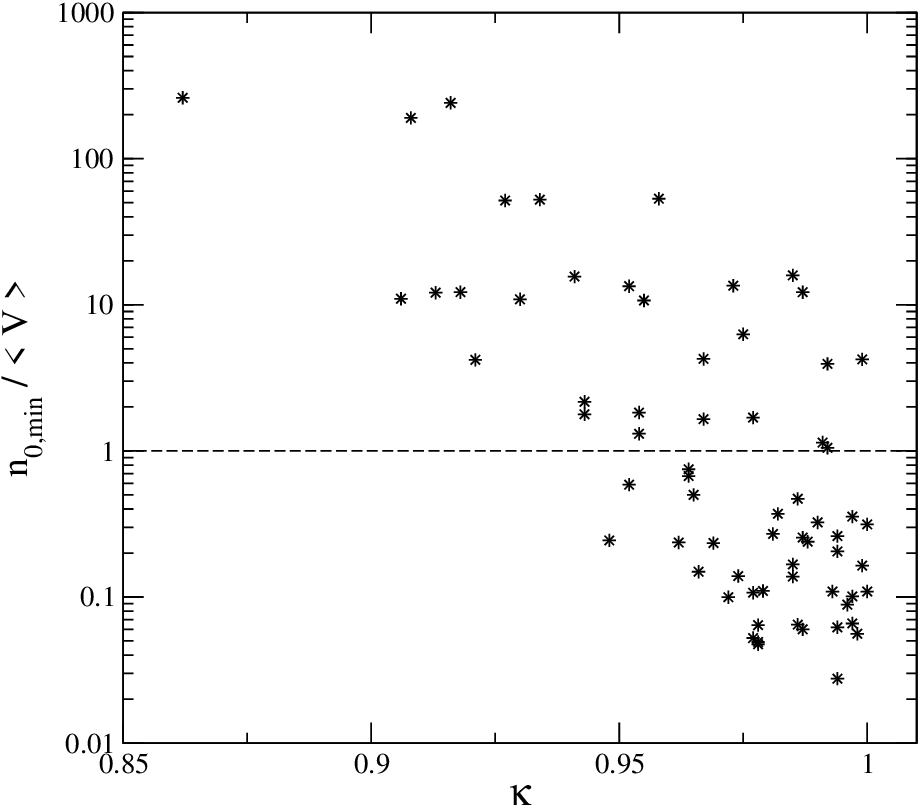}}
\caption{Fraction of daily volume needed to inject an exploitable arbitrage versus $\kappa$ (same data set) }\label{fig:n0mindivavgV_vs_kappa}
\end{figure}

The above discussion neglects the bid-ask spread $s$. It is of great importance in practice, as the impact of one trade is on average of the same order of magnitude as the spread \cite{BouchaudLimit5}. This means that $n_0$ must be large enough in order to make the knowledge of trader $1$ valuable. It is easy to convince oneself that it is enough to replace $n_0$ by $n_0'=I^{-1}[I(n_0)+\avg{s}]$ in the relevant equations, and multiply all the gains by $e^{\avg{s}/2}$. For example, the optimal number of shares that trader $1$ invests if trader $0$ has infinite capital is
\be
n_{1,s}^*=e^{-\avg{s}/\gamma}\frac{n_0}{(1+\gamma)^{1/\gamma}}
\ee
From this equation one sees that $n_0'=n_0\exp(-\avg{s}/\gamma)$. Since $\avg{s}/\gamma\sim10$ in practice, the minimal amount of shares needed to create an arbitrage, denoted by $n_{0,\min}$ is increased about 20,000 folds by the spread. The respective values of $\gamma$ and $\avg{s}$ are not independent, and can be measured in real markets for a given stock. In the language of \cite{BouchaudLimit3}, $\gamma=\avg{\log(n)}/R(1)$ where $\avg{\log (n)}$ is the average of the logarithm of transaction size and $R(1)$ is the response function after one time step. Using $\gamma$ and $\avg{s}$ measured in Paris Stock Exchange one finds that $n_{0,\min}\in[1.2 10^4,2.0 10^8]$, with median of  $1.2 10^5$ (Fig.\ \ref{fig:avgV_vs_n0min}), which is not unrealistic for very liquid stocks. Indeed, the fraction of $n_{0,\min}$ with respect to  the average daily volume of each stock ranges from $\simeq2\%$ to more than $100\%$, with a median of about $42\%$. Therefore, for most of the stocks, trader $0$ needs to trade less than two fifths of the daily average volume in one transaction in order to be leave an exploitable arbitrage; for 12 stocks ($18\%$), trading less than $10\%$ of the average daily volume suffices. It is unlikely that a single trade is larger than the average daily volume, hence, 29 stocks ($43\%$) do not allow {\em on average} a single large trade to be exploitable by a simple round-trip. Interestingly, stocks with average daily volumes smaller than about $10^5$ are all consistent from that point of view (Fig.\ \ref{fig:avgV_vs_n0min}). In addition, the stocks for which $n_{0,\min}<\avg{ V}$ all have a $\kappa>0.945$.

Therefore, the role of the spread is to increase considerably the minimum size of the trade, which in some cases remain within reasonable bounds. The mathematical discussions of the previous sections on $\log$ price impact functions are therefore still valid, provided that one replaces $n_0$ with $n_0'=I^{-1}[I(n_0)+s]$, which is equivalent to rescaling $n_0$ by $\exp(\avg{s}/\gamma)$. Therefore, the spread must be taken into account, but does not yield systematically consistent impact functions for some stocks with a high enough daily volume.

\section{Spread and feedback}

The question is whether the feedback and the spread make impact function systematically consistent. The stocks that are the most likely to become consistent are those whose $n_{0,\min}/\avg{ V}<1$ is large while having a strong feedback. According to  Fig.\ \ref{fig:n0mindivavgV_vs_kappa}, these properties are compatible.

Using for each stock $\avg{s}$, $\kappa$, and $\gamma$ from the data, we find that three additional stocks are made consistent by feedback on trader $0$'s market order side alone: the feedback limited to one side of the order book, even when the spread is taken into account, is insufficient. However, adding finally the feedback on both book sides makes consistent {\em all} the stocks, even in the case of infinite capital. Therefore, both the spread and the feedback are crucial ingredients of consistency at the smallest time scale.

\section{Conclusion}

 The paradox proposed in this paper provides an {\em a contrario} simple and necessary condition of consistency for price impact functions. And indeed, financial markets ensure consistent market price impact functions at the most microscopic dynamical level by two essential ingredients: the spread and the dynamics of the order book.


\bigskip

I am indebted to Jean-Philippe Bouchaud, Julien Kockelkoren and Michele Vettorazzo from CFM for their hospitality, measurements,  and comments. It is a pleasure to acknowledge discussions and comments from Doyne Farmer, Sam Howison, Matteo Marsili, Sorin Solomon and David Br\'ee.

\bibliographystyle{plain}
\bibliography{biblio}

\end{document}